\documentclass[floatfix,aps, prl, reprint, amsfonts, amssymb, amsmath, showpacs, a4paper, nofootinbib, superscriptaddress]{revtex4-1}

\usepackage{color}
\usepackage{graphicx}
\usepackage{braket}
\usepackage{microtype} 
\usepackage{setspace} 

\newcommand{\be}{\begin{equation}}
\newcommand{\ee}{\end{equation}}
\newcommand{\bea}{\begin{eqnarray}}
\newcommand{\eea}{\end{eqnarray}}
\newcommand{\nn}{\nonumber\\}

\usepackage[caption=false]{subfig}

\begin{document}

\title{Oscillation probabilities for a $\mathcal{PT}$-symmetric non-Hermitian two-state system}

\author{Jean Alexandre}
\email{jean.alexandre@kcl.ac.uk}
\affiliation{Department of Physics, King's College London, 
London WC2R 2LS, United Kingdom}

\author{Madeleine Dale}
\email{m.dale@stimulate-ejd.eu}
\affiliation{Universit\`{a} di Roma Tor Vergata, Dip.\ di Fisica,
Via della Ricerca Scientifica 1, 00133 Rome, Italy}
\affiliation{INFN, Sezione di Tor Vergata,
Via della Ricerca Scientifica 1, 00133 Rome, Italy}
\affiliation{Department of Physics, University of Cyprus, 1 Panepistimiou Street, 2109 Aglantzia, Nicosia, Cyprus}
\affiliation{Humboldt Universität zu Berlin, Institut für Physik {\&} IRIS Adlershof, Zum Gro{\ss}en Windkanal 6, 12489
Berlin, Germany}

\author{John Ellis}
\email{john.ellis@cern.ch}
\affiliation{Department of Physics, King's College London, 
London WC2R 2LS, United Kingdom}
\affiliation{Theoretical Physics Department, CERN,\\ CH-1211 Geneva 23, Switzerland}

\author{Robert Mason}
\email{mason@physik.rwth-aachen.de}
\affiliation{Department of Mathematical Sciences, University of Liverpool, Liverpool L69 7ZL, UK}
\affiliation{Institute for Theoretical Particle Physics and Cosmology, RWTH Aachen University, 52056 Aachen, Germany}

\author{Peter Millington}
\email{Corresponding{\color{white}~}author:{\color{white}~}peter.millington@manchester.ac.uk}
\affiliation{Department of Physics and Astronomy, University of Manchester, Manchester M13 9PL, United Kingdom}

\date{17 February 2026}

\begin{abstract}
    
    There is growing interest in viable quantum theories with $\mathcal{PT}$-symmetric non-Hermitian Hamiltonians, but a formulation of transition matrix elements consistent with positivity and perturbative unitarity has so far proved elusive. This Letter provides such a formulation, which relies crucially on the ability to span the state space in such a way that the interaction and energy eigenstates are orthonormal with respect to the same positive-definite inner product. 
    We apply this non-Hermitian approach to two-neutrino flavour oscillations, and show how it can accommodate the seesaw mechanism.\\
~~\\\noindent
KCL-PH-TH/2023-17, CERN-TH-2023-032, LTH 1334
\end{abstract}

\maketitle


Non-Hermitian quantum field theories have applications in many areas of physics and have attracted wide interest: see Ref.~\cite{Bender:2005tb} and the numerous references therein. Such theories offer the possibility of broadening the Hermitian framework of the Standard Model of particle physics and constructing novel viable extensions of it. With this motivation in mind, there has been significant progress in the analysis of non-Hermitian quantum field theories, including the formulation of spontaneous symmetry breaking, the
Goldstone theorem and the Englert-Brout-Higgs mechanism~\cite{Alexandre:2018uol, Mannheim:2018dur, Fring:2019hue}. 

The dynamics of particle mixing are well studied for Hermitian quantum theories. They have been applied with great success to the phenomena of flavour oscillations, such as meson mixing 
and neutrino oscillations (see, e.g., Ref.~\cite{ParticleDataGroup:2024cfk}). These oscillations arise due to the misalignment of the diagonal bases of the interaction terms and the mass terms in the corresponding quantum field theory. In a misaligned situation, an interaction eigenstate can be decomposed in terms of a superposition of energy eigenstates. 
Since each of the energy eigenstates evolves with a different phase, the interaction eigenstate is not stationary. 
There is then a non-zero probability of measuring a different interaction eigenstate at some later time.

However, a satisfactory description of oscillation phenomena in systems with non-Hermitian but $\mathcal{PT}$-symmetric mixing has so far proved elusive. This is despite the fact that the viability of non-Hermitian quantum theories is well established in the presence of some antilinear symmetry 
of the Hamiltonian $\hat{H}$.  Examples include $\mathcal{PT}$ (parity-time-reversal)-symmetric quantum theories~\cite{Bender:2005tb}, wherein $[\hat{H},\mathcal{PT}]=0$, 
and the more general class of pseudo-Hermitian quantum theories~\cite{Mostafazadeh:2001jk}.  The problem lies in the following observation: 
Whilst unitarity is guaranteed in, e.g., $\mathcal{PT}$-symmetric theories, due to the existence of an additional 
discrete symmetry of the Hamiltonian~\cite{Bender:2002vv}, existing analyses have arrived at individual transition probabilities 
that can be negative or larger than unity~\cite{AEMB, Ohlsson:2019noy}.

Motivated by the desire to construct viable non-Hermitian extensions of the Standard Model, in this Letter we resolve this longstanding theoretical issue, highlighting the differences between oscillation probabilities in Hermitian and non-Hermitian theories, and laying the basis for phenomenological analyses of the latter.

We consider first a simple and well-studied quantum field-theoretic model 
comprising two complex scalar fields $\phi_1$ and $\phi_2$ (introduced in Ref.~\cite{Alexandre:2017foi}) that can be arranged in a complex doublet 
$\Phi=(\phi_1,\phi_2)$, which mix via a non-Hermitian mass matrix $M\neq M^{\dag}$. However, our analysis holds for any two-state system with a non-Hermitian but $\mathcal{PT}$-symmetric Hamiltonian, as considered, e.g., in Ref.~\cite{Ohlsson:2019noy}. 

The Lagrangian density for the scalar field theory is
\be
\label{nonHL}
\mathcal{L} = \partial_{\alpha}\tilde{\Phi}^{\dag}\partial^{\alpha}\Phi-\tilde{\Phi}^{\dag}M^2\Phi \, ,
\ee
where $\partial_{\alpha}$ is a spacetime derivative and the squared mass matrix is
\be
M^2 = \begin{bmatrix} m_1^2 && \mu^2 \\ -\mu^2 && m_2^2 \end{bmatrix} \neq (M^2)^{\dag} \, .
\ee
The formulation in Eq.~\eqref{nonHL} of the dynamics in terms of the tilde-conjugate doublet $\tilde{\Phi}^{\dag}\neq\Phi^{\dag}$ (where $\dag$ denotes Hermitian conjugation), first introduced in Ref.~\cite{AEMB}, is necessary for the mutual consistency of the Euler--Lagrange equations obtained directly by varying this Lagrangian.

The squared mass eigenvalues
\be
m_\pm^2=\frac{1}{2}(m_1^2+m_2^2)\pm\frac{1}{2}\sqrt{(m_1^2-m_2^2)^2-4\mu^4}
\ee
are real, so long as the argument of the square root is positive,  and the corresponding eigenvectors are
\begin{subequations}
\bea
\label{e1e2}
\mathbf{e}_+&=&N\begin{bmatrix} \eta \\ -1+\sqrt{1-\eta^2} \end{bmatrix} \, ,\\
\mathbf{e}_-&=&N\begin{bmatrix} -1+\sqrt{1-\eta^2} \\ \eta \end{bmatrix}\,,
\eea
\end{subequations}
where the normalisation factor $N$ is defined below and
\be
\eta \equiv \frac{2\mu^2}{|m_1^2-m_2^2|} \, .
\ee

The parameter $\eta$ must be less than or equal to unity for the eigenvalues to be real. At $\eta=1$, 
the eigenvalues coalesce, corresponding to an exceptional point at which the mass matrix becomes defective. 
Such exceptional points are novel features of non-Hermitian quantum theories and we will see that, 
in the context of flavour oscillations, the exceptional point is that at which the transition probabilities saturate 
for finite values of the Lagrangian parameters, in stark contrast to the Hermitian case.

Since the squared mass matrix is not Hermitian, it is diagonalised by a similarity (rather than orthogonal) transformation, and its eigenvectors are not orthogonal with respect to the usual Hermitian scalar product (the Dirac inner product).  Nevertheless, there exists in the regime where the eigenvalues are real an orthogonal inner product, which has been described at length in the existing literature, both in the case of non-Hermitian quantum mechanics (see, e.g, Refs.~\cite{Mostafazadeh:2001jk, Bender:2002vv, Bender:2005tb, Mannheim:2017apd}) and non-Hermitian quantum field theory (see Ref.~\cite{AEMB} for the present scalar theory and Ref.~\cite{Alexandre:2022uns} for a related Dirac fermion theory). Positive norms are obtained with respect to the so-called $\mathcal{C}'\mathcal{PT}$ inner product, where the transformation $\mathcal{C}'$ (not to be confused with charge conjugation in the case of quantum field theory, see Ref.~\cite{AEMB}) is an additional discrete symmetry of the Hamiltonian, i.e., $[\hat{H},\mathcal{C}']=0$, and this symmetry ensures unitarity~\cite{Bender:2002vv}.
Finally, we note also that a scalar field theory may be considered as the continuous limit of a discrete system of spins, for which non-Hermitian Hamiltonians can be defined, see, e.g., \cite{Kattel:2023ras}. In this context, one should also be able to construct a positive norm, since this is required for a consistent description of the $\mathcal{PT}$-symmetric phase.

The eigenvectors $\mathbf{e}_+$ and $\mathbf{e}_-$ are orthogonal with respect to the $\mathcal{PT}$ inner product
\bea
\mathbf{e}_{\pm}^{\ddag}\mathbf{e}_{\pm} = \mathbf{e}_{\pm}^{\dag}P \mathbf{e}_{\pm}=\pm1\, ,\qquad
\mathbf{e}^{\ddag}_{\pm} \mathbf{e}_{\mp} = 0\, ,
\eea
where $\ddag\equiv \mathcal{PT}\circ \mathsf{T}$, with $\mathsf{T}$ denoting matrix transposition. Since we work hereafter with a $2\times 2$ matrix representation, time-reversal simply effects complex conjugation. We have fixed the normalisation~\cite{AEMB}
\be
N = \left[2\left(\eta^2-1+\sqrt{1-\eta^2}\right)\right]^{-1/2}\,.
\ee
This normalisation diverges in the Hermitian limit $\eta\to 0$, but the normalised eigenvectors themselves remain well defined. In addition, we have introduced the parity matrix
\be
P = \begin{bmatrix} 1 & 0 \\ 0 & -1\end{bmatrix}\,,\qquad P^2 = \mathbb{I}\, ,
\ee
which satisfies $P M^2 P = (M^2)^{\dag}$. The eigenvector $\mathbf{e}_-$ has negative $\mathcal{PT}$ norm, but its $\mathcal{C}'\mathcal{PT}$ norm is positive:
\bea
    \mathbf{e}^{\S}_{\pm}\mathbf{e}_{\pm} = \mathbf{e}_{\pm}^{\dag}C' P \mathbf{e}_{\pm}=1\, ,\qquad 
\mathbf{e}^{\S}_{\pm}\mathbf{e}_{\mp} = 0\, ,
\eea
where $\S\equiv \mathcal{C'PT}\circ \mathsf{T}$ and
\be
C' =  \frac{1}{\sqrt{1-\eta^2}}\begin{bmatrix} 1 & -\eta \\ \eta & -1\end{bmatrix}\,, \qquad (C')^2 = \mathbb{I}\;,
\ee
and we have used $(C'\cdot P)^{\mathsf{T}}=C'\cdot P$ (see Ref.~\cite{AEMB}).

In order to calculate the transition and survival probabilities of the flavour states, we introduce a two-dimensional state space spanned by the eigenvectors $\mathbf{e}_+$ and $\mathbf{e}_-$. These are related to the flavour kets $\left|\phi_{i,\vec{p}}(t,\vec{x})\right>$ by the similarity transformation that diagonalises the squared mass matrix, and we obtain
\begin{subequations}
\label{flavourkets}
\bea
\ket{\phi_{1,\vec{p}}(x)}&=&\cosh(\theta)\,\xi_{+,\vec{p}}(x)\,\mathbf{e}_++\sinh(\theta)\,\xi_{-,\vec{p}}(x)\,\mathbf{e}_- \, ,\nonumber\\ \\
\ket{\phi_{2,\vec{p}}(x)}&=&\cosh(\theta)\,\xi_{-,\vec{p}}(x)\,\mathbf{e}_-+\sinh(\theta)\,\xi_{+,\vec{p}}(x)\,\mathbf{e}_+ \, ,\nonumber\\
\eea
\end{subequations}
where $\theta=\frac{1}{2}\mbox{arctanh}(\eta)$, such that
\begin{subequations}
\bea
\cosh(\theta)&=&\frac{1}{\sqrt2}\left(1+\frac{1}{\sqrt{1-\eta^2}}\right)^{1/2}\,,\\
\sinh(\theta)&=&\frac{1}{\sqrt2}\frac{\eta}{\sqrt{1-\eta^2}}\left(1+\frac{1}{\sqrt{1-\eta^2}}\right)^{-1/2}\,.\quad
\eea
\end{subequations}
The eigenfunctions $\xi_{\pm,\vec{p}}(x)$ satisfy the classical equations of motion $(\Box+m_{\pm}^2)\xi_{\pm}=0$ ($\Box\equiv \partial_{\alpha}\partial^{\alpha}$), with solutions
\bea
\xi_{\pm,\vec{p}}(x)&=&\exp(i\omega_{\pm}t+i\vec p\cdot\vec x) \, ,
\eea
where \smash{$\omega_{\pm}=\sqrt{\vec p^{\,2}+m_{\pm}^2}$}. Hereafter, for simplicity, we will consider only the zero-momentum modes with $\vec{p}=\vec{0}$. At $t=0$, the flavour states then reduce to
\be
\ket{\phi_1(0)} = \begin{bmatrix}1 \\ 0\end{bmatrix} \, ,\qquad
\ket{\phi_2(0)} = \begin{bmatrix}0 \\ 1\end{bmatrix} \, ,
\ee
as we would expect.

The flavour-conjugate states (see Ref.~\cite{AEMB}) can be expressed in the form
\begin{subequations}
\label{tildebras}
\bea
\bra{\tilde{\phi}_1(t)}&=&\cosh(\theta)\,\xi_+^{\ast}(t)\,\mathbf{e}_+^{\S}-\sinh(\theta)\,\xi_-^{\ast}(t)\,\mathbf{e}_-^{\S} \, ,\ \\
\bra{\tilde{\phi}_2(t)}&=&\cosh(\theta)\,\xi_-^{\ast}(t)\,\mathbf{e}_-^{\S}-\sinh(\theta)\,\xi_+^{\ast}(t)\,\mathbf{e}_+^{\S} \, .\
\eea
\end{subequations}
We emphasise the change of sign $\sinh(\theta)\to-\sinh(\theta)$ relative to the states in Eq.~\eqref{flavourkets}. These states satisfy
\be
\braket{\tilde{\phi}_i(t)|\phi_j(t)} = \delta_{ij} \, ,
\ee
and, at $t=0$, the conjugate states reduce to
\be
\bra{\tilde{\phi}_1(0)} = \begin{bmatrix}1 & 0\end{bmatrix} \, ,\qquad
\bra{\tilde{\phi}_2(0)} = \begin{bmatrix}0 & 1\end{bmatrix} \, .
\ee
The latter are, in fact, the Hermitian-conjugate flavour states, as identified in Ref.~\cite{AEMB} [see  Eq.~(50) therein]. Thus, the flavour states are orthogonal with respect to the Dirac inner product (constructed via Hermitian conjugation) at $t=0$, but cease to be so for any $t\neq 0$. As a result (see the Supplemental Material), attempts to construct transition probabilities using the Dirac inner product necessarily lead to the violation of time-translation invariance.

An important observation is that the conjugate states in Eq.~\eqref{tildebras} do not coincide with the $\mathcal{C}'\mathcal{P}\mathcal{T}$-conjugates of the flavour states, which are instead given by
\begin{subequations}
\label{CpPTbras}
\bea
\bra{\phi_1^{\mathcal{C}'\mathcal{PT}}(t)}&=&\cosh(\theta)\,\xi_+^{\ast}(t)\,\mathbf{e}_+^{\S}+\sinh(\theta)\,\xi_-^{\ast}(t)\,\mathbf{e}_-^{\S} \, ,\nonumber \\ \\
\bra{\phi_2^{\mathcal{C}'\mathcal{PT}}(t)}&=&\cosh(\theta)\,\xi_-^{\ast}(t)\,\mathbf{e}_-^{\S}+\sinh(\theta)\,\xi_+^{\ast}(t)\,\mathbf{e}_+^{\S}\,.\nonumber \\
\eea
\end{subequations}
These do not, however, provide an orthogonal basis with respect to $\mathcal{C}'\mathcal{P}\mathcal{T}$:
\be
\braket{\phi_i^{\mathcal{C}'\mathcal{PT}}(t)|\phi_j(t)} = \begin{cases}\cosh(2\theta)\,,\qquad &i = j\\
\sinh(2\theta)\,,\qquad &i \neq j\,.
\end{cases}
\ee
At $t=0$, the states in Eq.~\eqref{CpPTbras} reduce to
\begin{subequations}
\bea
\bra{\phi_1^{\mathcal{C}'\mathcal{PT}}(0)}&=&\frac{1}{\sqrt{1-\eta^2}}\begin{bmatrix}1 & \eta\end{bmatrix}\, ,\\
\bra{\phi_2^{\mathcal{C}'\mathcal{PT}}(0)}&=&\frac{1}{\sqrt{1-\eta^2}}\begin{bmatrix}\eta & 1\end{bmatrix} \, ,
\eea
\end{subequations}
and we see that the $\mathcal{C}'\mathcal{PT}$-conjugates of the flavour states do not have a direct interpretation as flavour states.

However, the choice of basis $\{\ket{\phi_1},\ket{\phi_2}\}$ is not unique, and care must be taken to avoid an unsuitable choice that yields unphysical results (viz.~the negative or unbounded ``probabilities'' mentioned earlier). In what follows, we resolve this longstanding issue by identifying the appropriate choice of basis.

Since the Hamiltonian is $\mathcal{C}'$-symmetric (i.e., $C^{\prime\mathsf{T}} M^2 C^{\prime\mathsf{T}}=M^2$), we can span the state space with $\{\ket{\phi_1},\ket{\phi_2^{\mathcal{C}'}}\}$ (or, equivalently, $\{\ket{\phi_1^{\mathcal{C}'}},\ket{\phi_2}\}$). Remarkably, this choice allows us to construct an orthonormal \emph{flavour} basis with respect to 
$\mathcal{C}'\mathcal{PT}$, with
\begin{subequations}
\begin{gather}
\braket{\phi_1^{\mathcal{C}'\mathcal{PT}}(t)|\phi_1(t)} = 1 \, ,\quad
\braket{\phi_2^{\mathcal{PT}}(t)|\phi_2^{\mathcal{C}'}(t)} = 1 \, ,\\
\braket{\phi_1^{\mathcal{C}'\mathcal{PT}}(t)|\phi_2^{\mathcal{C}'}(t)} = 0 \, ,\quad 
\braket{\phi_2^{\mathcal{PT}}(t)|\phi_1(t)} = 0\,,
\end{gather}
\end{subequations}
where we have adjusted the normalisations of all of the flavour states by a factor of $\sqrt{\mathrm{sech}(2\theta)}$. Crucially, with this choice of basis, the \emph{flavour} and \emph{mass} eigenstates are orthonormal with respect to the same positive-definite inner product. While all the inner products are with respect to $\mathcal{C}'\mathcal{PT}$, the choice to span the flavour space by $\ket{\phi_1}$ and the $\mathcal{C}'$ conjugate of $\ket{\phi_2}$ means that the inner product between different flavour states, in fact, reduces to the $\mathcal{PT}$ inner product, since, e.g., $\braket{\phi_2^{\mathcal{PT}}(t)|\phi_1(t)}=(\ket{\phi_2^{\mathcal{C'}}(t)})^{\S}\ket{\phi_1(t)}$. Notice that the norms of both the flavour states are nevertheless with respect to $\mathcal{C}'\mathcal{PT}$.

Spanning the state space in this way, our initial density operators are
\begin{subequations}
\bea
\hat{\rho}_1(t_0)&=&\ket{\phi_1(t_0)}\bra{\phi_1^{\mathcal{C}'\mathcal{PT}}(t_0)} \, ,\\
\hat{\rho}_2(t_0)&=&\ket{\phi_2^{\mathcal{C}'}(t_0)}\bra{\phi_2^{\mathcal{PT}}(t_0)} \, ,
\eea
\end{subequations}
and the final-state projection operators are
\begin{subequations}
\bea
\hat{\pi}_1(t)&=&\ket{\phi_1(t)}\bra{\phi_1^{\mathcal{C}'\mathcal{PT}}(t)} \, ,\\
\hat{\pi}_2(t)&=&\ket{\phi_2^{\mathcal{C}'}(t)}\bra{\phi_2^{\mathcal{PT}}(t)} \, .
\eea
\end{subequations}
The transition and survival probabilities are calculated as $\mathbb{P}_{i\to j}(t,t_0)=\mathrm{tr}\hat{\rho}_i(t_0)\hat{\pi}_j(t)$, and we obtain
\begin{subequations}
\bea
\mathbb{P}_{1(2)\to1(2)}(t,t_0)&=&1-\eta^2\sin^2\left[\Delta \omega \Delta t/2\right] \, ,\\
\mathbb{P}_{1(2)\to2(1)}(t,t_0)&=&\eta^2\sin^2\left[\Delta \omega \Delta t/2\right] \, ,
\eea
\end{subequations}
where $\Delta\omega\equiv \omega_+-\omega_-$ and $\Delta t\equiv t-t_0$. These probabilities are consistent with positivity, unitarity, perturbative unitarity (in that they are finite for all $\eta\in[0,1]$) and respect time-translation invariance (cf.~the Supplemental Material). 

\begin{figure}
    \centering
    \subfloat[Oscillation probabilities for the Hermitian (topmost two surfaces, \smash{$\mathbb{P}^{\rm Herm}_{i\to j}$}) and non-Hermitian (bottommost two surfaces, \smash{$\mathbb{P}^{\mathcal{PT}}_{i\to j}$}) models as a function of $\eta$ and the phase $\vartheta\equiv\Delta \omega \Delta t/2$. The legend is ordered from topmost to bottommost surfaces.]{
    \centering
    \includegraphics[width=0.45\textwidth]{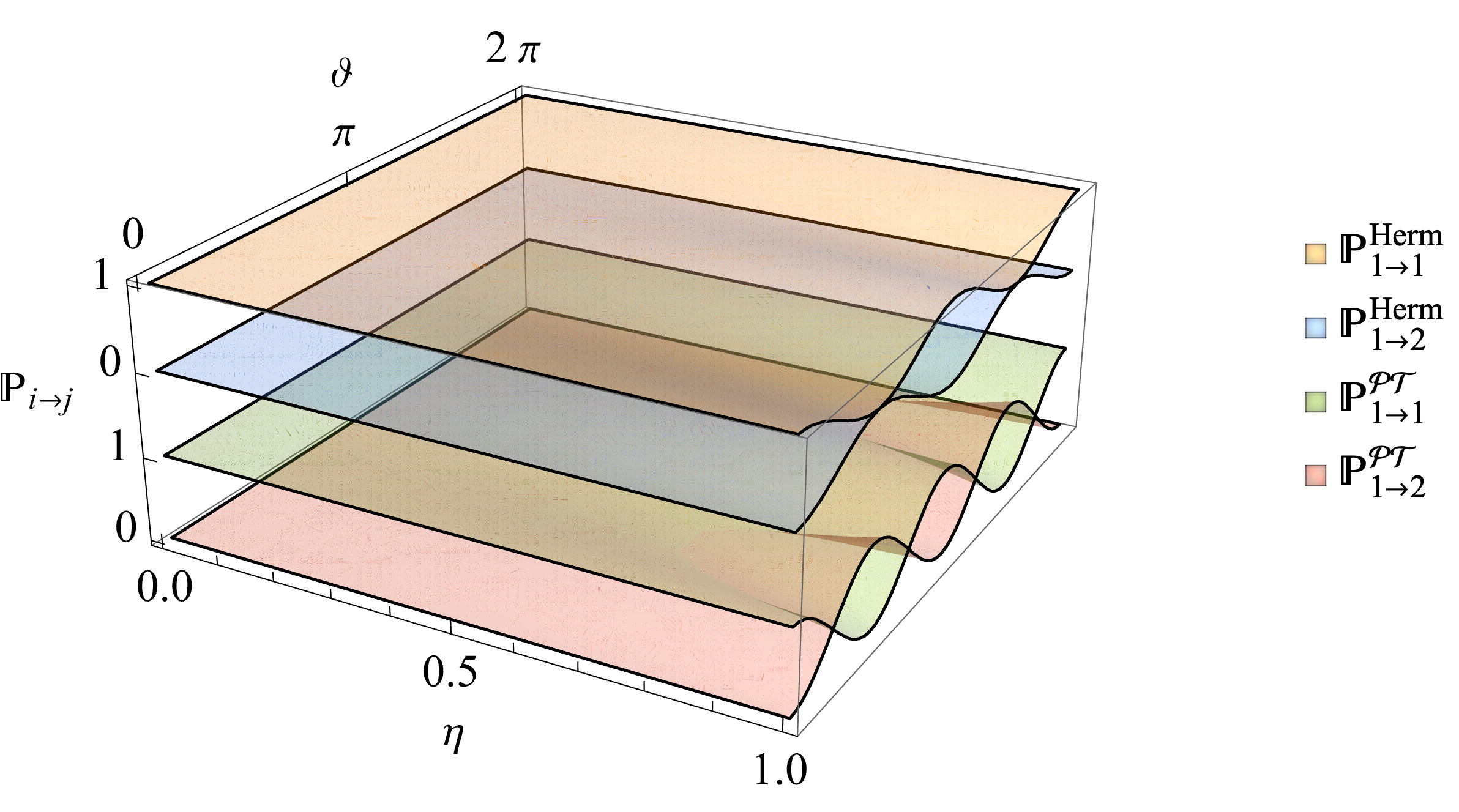}
    \label{probabilitiesfig}
    }\\
    \subfloat[Squared eigenmasses of the Hermitian ($m^2_{\pm\rm{Herm}}$) and non-Hermitian ($m^2_{\pm\mathcal{PT}}$) models divided by $m_1^2+m_2^2$ versus $\eta$ for $(m_1^2-m_2^2)/(m_1^2+m_2^2)=0.5$.]{
    \centering
    \includegraphics[width=0.38\textwidth]{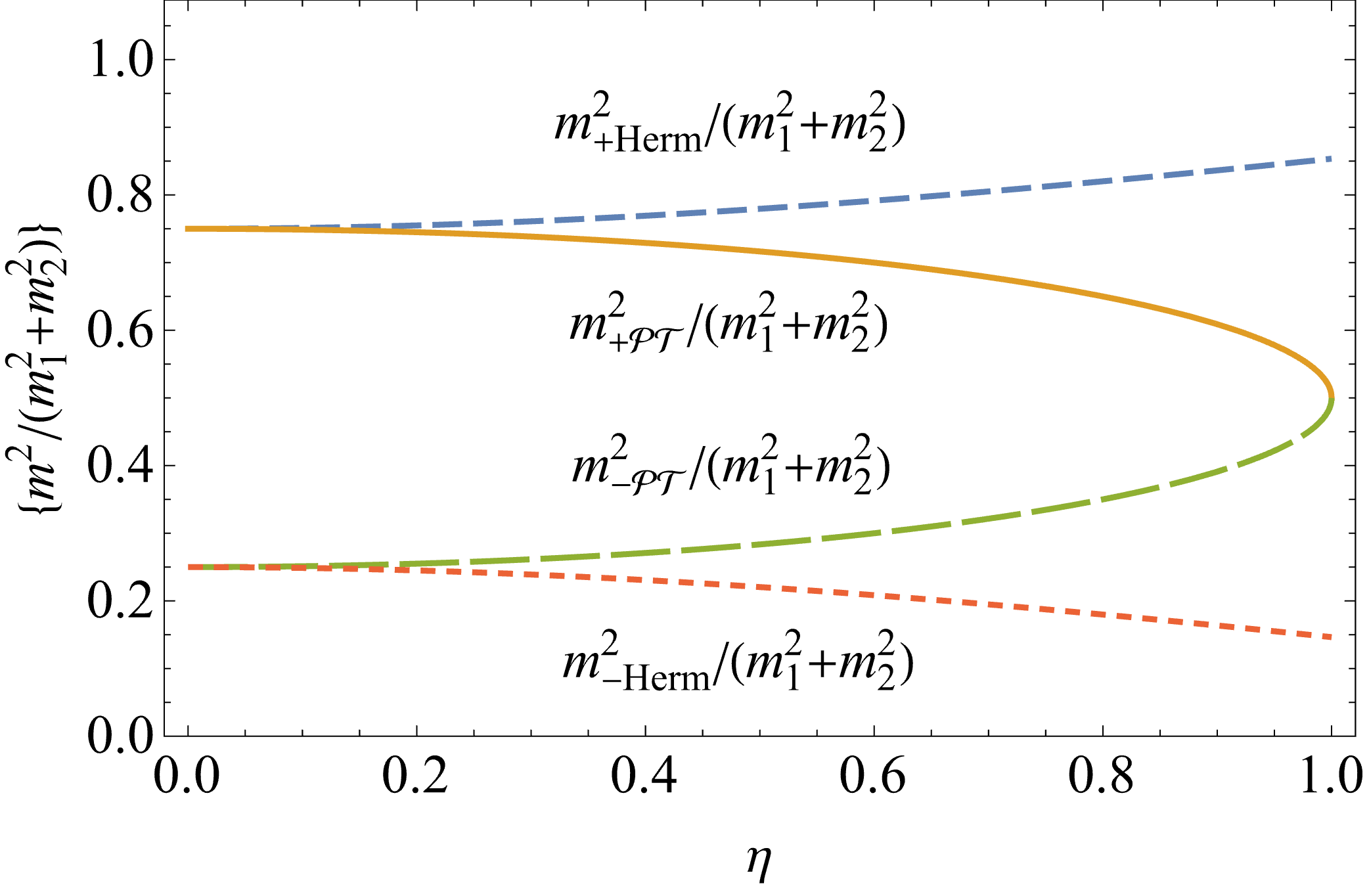}
    \label{massesfig}
    }
    \caption{Comparison of the transition and survival probabilities (a) and squared eigenmasses (b) for the Hermitian and non-Hermitian models as a function of the parameter $\eta$. For the Hermitian case, the mass eigenvalues diverge for large $\eta$, with the lower eigenvalue crossing zero and becoming negative at a value of $\eta^2=(m_1^2+m_2^2)^2/(m_1^2-m_2^2)^2-1$. This occurs before the transition and survival probabilities saturate. For the non-Hermitian case, the mass eigenvalues merge at the exceptional point $\eta=1$, at which the probabilities saturate.}
    \label{comparisonfig}
\end{figure}

We note, however, that these are not the analytic continuations via $\mu^4\to-\mu^4$ of the corresponding probabilities for the model with Hermitian mass mixing given by taking
\be
    M^2 \rightarrow M^2_{\rm Herm} = \begin{bmatrix} m_1^2 &\mu^2 \\ \mu^2 & m_2^2\end{bmatrix} \, ,
    \qquad \tilde{\Phi}^{\dag} \rightarrow \Phi^{\dag}\,,
\ee
in Eq.~\eqref{nonHL}. The transition probabilities for this Hermitian model are given by
\bea
\label{HermProb}
&&\mathbb{P}^{\rm Herm}_{1(2)\to2(1)}(t,t_0) =  
\frac{\eta^2}{1+\eta^2}\sin^2\left[\Delta \omega \Delta t/2\right] \, .
\eea
Whereas the Hermitian case saturates for $\eta\to\pm \infty$, the non-Hermitian probabilities saturate at the exceptional point 
$\eta\to \pm 1$ ($\mu^2=\pm (m_1^2-m_2^2)/2$) (see Fig.~\ref{probabilitiesfig}). Moreover, the masses become degenerate at this exceptional point, but they diverge in the Hermitian case, with the lower squared mass becoming negative for sufficiently large mixing, signalling a tachyonic instability (see Fig.~\ref{massesfig}). Note that the analytic continuation $\mu^4\to-\mu^4$ of the Hermitian result in Eq.~\eqref{HermProb} (as reported in Ref.~\cite{AEMB}) would be negative, with a modulus exceeding unity for $\eta>1/\sqrt{2}$.

All our key results are illustrated in Fig.~\ref{comparisonfig}, where in the upper panel we see explicitly that the survival and transition probabilities have very different behaviours in terms of $\eta$ and the phase $\vartheta = \Delta \omega \Delta t/2$ between the Hermitian and non-Hermitian models, and in the lower panel we see differences in the mass eigenstates as functions of $\eta$ for $(m_1^2-m_2^2)/(m_1^2+m_2^2)=0.5$. These results are in principle directly applicable to the analysis of meson mixing, specifically in the $K^0 - \bar K^0$, $D^0 - \bar D^0$ and $B^0_{d,s} - \bar B^0_{d,s}$ systems (see, e.g., Sec.~12 of Ref.~\cite{ParticleDataGroup:2024cfk}). However, phenomenological studies of these systems lie beyond the scope of this Letter. 

Instead, we consider here the example of two-neutrino mixing, and the consistency of our approach with the seesaw mechanism.
The potential relevance of $\mathcal{PT}$-symmetric extensions of the Standard Model Lagrangian to neutrino oscillations was first identified in Ref.~\cite{Jones-Smith:2009qeu} (see also the later works~\cite{Alexandre:2015kra, Mishra:2018aej, Rodionov:2022hyq}).

We consider two fermion flavours $\nu_i$, each composed of one left-handed Weyl fermion $\chi_i$ and one right-handed Weyl fermion $\zeta_i$, 
\be
\nu_i=\begin{pmatrix} \chi_i \\ \zeta_i \end{pmatrix}\,.
\ee
We assume that these flavours are coupled with the non-Hermitian mass term
\bea\label{massterm2}
\mathcal{L}\supset -\begin{pmatrix}\bar{\nu}_1 & \bar{\nu}_2 \end{pmatrix}
\begin{pmatrix}
     M_1 & \mu J \\  
    -\mu J & M_2 
\end{pmatrix}   
\begin{pmatrix} \nu_1 \\ \nu_2 \end{pmatrix}\,,
\eea
where $M_i$ and $J$ are $2\times2$ Hermitian matrices that act on the chiral components $\chi_i,\zeta_i$. 
We choose for $M_1$ and $M_2$ a structure allowing a seesaw mechanism for each individual flavour:
\be
M_i=\begin{pmatrix} b_i & m_i \\ m_i & b_i' \end{pmatrix}\,,
\ee
where $b_i, b_i'\gg m_i$. We will see below that the consistent choice is to take $b_1=b_2$, $b_1'=b_2'$, and $J$ of the form
\be\label{J}
J=\begin{pmatrix} 0 & 1 \\ 1 & 0 \end{pmatrix}\,.
\ee    
We use a similarity transformation
\be
\label{eq:S_def}
S=\begin{pmatrix} c & s \\ s & c \end{pmatrix}\,,
\ee
where $c\equiv\cosh\theta$, $s\equiv\sinh\theta$, to diagonalise the flavour structure as
\be\label{diagonal}
\begin{pmatrix}
     M_1 & \mu J \\  
    -\mu J & M_2 
\end{pmatrix}  
=S\begin{pmatrix} M_- & 0 \\ 0 & M_+ \end{pmatrix}S^{-1}\,,
\ee
where $M_\pm$ are the mass matrices for the mass eigenstates
\be
\begin{pmatrix} \nu_- \\ \nu_+ \end{pmatrix}=
S^{-1}\begin{pmatrix} \nu_1 \\ \nu_2 \end{pmatrix}\,.
\ee
From the identity (\ref{diagonal}) we find 
\bea
M_-=\frac{c^2M_1+s^2M_2}{c^2+s^2}\,,\quad 
M_+=\frac{s^2M_1+c^2M_2}{c^2+s^2}\,,
\eea
together with the constraint
\be
cs(M_2-M_1)=\mu(c^2+s^2)J\,.
\ee
Thus, $M_2-M_1$ should be proportional to $J$, which is possible with $b_1=b_2$, $b_1'=b_2'$,
and the choice (\ref{J}) for $J$. The hyperbolic mixing angle $\theta$ should then satisfy
\be
cs(m_2-m_1)=\mu(c^2+s^2)\,,
\ee
implying that
\be
\tanh(2\theta)=\frac{2\mu}{m_2-m_1}\,,
\ee
cf.~Ref.~\cite{Alexandre:2020wki}. The above expression is consistent for $2|\mu|<|m_1-m_2|$ only, corresponding to the $\mathcal{PT}$ regime of real eigenmasses [see Eq.~\eqref{eigenmasses}].
From the above results, and if we denote $b\equiv b_1=b_2$, $b'\equiv b_1'=b_2'$, we finally find 
\be
M_\pm=\begin{pmatrix} b & m_\pm \\ m_\pm & b' \end{pmatrix}\,,
\ee
where
\bea\label{eigenmasses}
m_\pm&=&\frac{m_1+m_2}{2}\pm\frac{m_1-m_2}{2}\sqrt{1-\tanh^2(2\theta)}\nn
&=&\frac{m_1+m_2}{2}\pm\frac{1}{2}\sqrt{(m_1-m_2)^2-4\mu^2}\,.
\eea
The latter eigenmasses coincide with those obtained in the limit of vanishing Majorana mass terms $b,b'$ 
and agree with those obtained in Ref.~\cite{Alexandre:2020wki}, where no Majorana terms were considered. 
The mass matrix~\eqref{eigenmasses} is identical to that obtained in the Hermitian case involving Majorana 
masses,\footnote{We note that in Ref.~\cite{Yoon:2000fc} the flavour and chiral components are arranged differently, 
leading to a different tensor product for the mass matrix, but which leads to the same eigenmasses.} when changing the sign $\mu^2\to-\mu^2$ \cite{Yoon:2000fc}.
Thus, we see that the structure of $M_\pm$ is compatible with realising the seesaw mechanism also in this non-Hermitian theory.

Turning now to the charged-current interactions for the first two generations of charged leptons and neutrinos, these take the form 
\begin{align}
 -\mathcal{L}_{\rm CC}&=\frac{g}{\sqrt{2}}\begin{pmatrix} \bar{e},\bar{\mu}\end{pmatrix}\gamma^{\mu}W^+_\mu V\begin{pmatrix} \nu_- \\ \nu_+\end{pmatrix}\nonumber\\
 &\qquad +\frac{g}{\sqrt{2}}\begin{pmatrix} \bar{\nu}_-,\bar{\nu}_+\end{pmatrix}V^{-1}\gamma^{\mu}W^-_\mu \begin{pmatrix} e \\ \mu\end{pmatrix}\,,
\end{align}
where
\begin{equation}
V_{ij}=F_{\ell,ii}U_{\ell,ik}^{\dag}S_{kj}F_{\nu,jj}\,.
\end{equation}
The $F_{\ell}$ and $F_{\nu}$ are diagonal matrices of charged-lepton and neutrino phases, respectively, and the unitary matrix $U_{\ell}$ is that appearing in the diagonalisation of the charged-lepton mass matrix (see, e.g., Sec.~14 of Ref.~\cite{ParticleDataGroup:2024cfk}). Since $S$ is not a unitary matrix, the matrix $V$ is also not unitary, except in the Hermitian limit.

A general $n\times n$ matrix $V$ is parametrised by $2n^2$ real numbers.  Taking the neutrinos to be of Dirac type, we can remove $2n-1$ phases, as in the usual Hermitian case. Unitarity would normally provide an additional $n^2$ constraints, leaving us with a matrix parametrised by $(n-1)^2$ real numbers, divided into $(n-1)/2$ angles and $(n-1)(n-2)/2$ phases. For $n=2$, this would leave us with one mixing angle. In the non-Hermitian case, however, we instead have constraints from the intertwining relation $VP = PV^{-1}$ ($P=P^{-1}$), which ensures the pseudo-Hermiticity of the Lagrangian. For $n=2$, this gives 3 constraints, leaving $2\times 2^2-(2\times 2-1)-3=2$ real numbers, i.e., one mixing angle and one phase. Thus, the most general form of the matrix $V$ is 
\begin{equation}
    \label{eq:Vmatrix}
    V=\begin{pmatrix} \cosh\theta & e^{i\varphi}\sinh\theta \\ e^{-i\varphi}\sinh \theta & \cosh\theta \end{pmatrix}\,.
\end{equation}
The matrix $V$ is Hermitian, as expected, and we note the additional phase in the mixing matrix compared to the matrix $S$ in Eq.~\eqref{eq:S_def} that diagonalised the mass matrix considered earlier.

We now consider the probability for the transition $\nu_{\mu}\to\nu_e$ between weak eigenstates. We have that
\begin{subequations}
\begin{align}
    \ket{\nu_{\mu}(t_0)}&=(V^{-1})_{\mu k}e^{i\omega_k t_0}\ket{\nu_k(0)}\,,\\
    \bra{
    \tilde{\nu}_{e}(t)}&=\bra{\nu_j(0)}e^{-i\omega_j t}V_{je}\,,
\end{align}
\end{subequations}
where $j,k=+,-$. Notice that $\bra{\tilde{\nu}_{e}(0)}$ is the flavour-conjugate state of $\ket{\nu_{e}(0)}$. To construct the transition amplitudes, we must consider instead the $\mathcal{P}\mathcal{T}$-conjugate state
\begin{equation}
    \bra{\nu_e^{\mathcal{PT}}(t)}=\bra{\nu_j^{\mathcal{PT}}(0)}e^{-i\omega_j t}(V^{-1})_{je}^*\,.
\end{equation}
We then have the amplitude
\begin{align}
    \label{eq:amp1}
    \braket{\nu_e^{\mathcal{PT}}(t)|\nu_{\mu}(t_0)}&={\rm sech} 2\theta\,e^{-i\omega_j t+i\omega_k t_0}\nonumber\\&\phantom{=} \quad\times P_{kj}(V^{-1})_{j e}^*(V^{-1})_{\mu k}\,,
\end{align}
and the conjugate amplitude is
\begin{align}
    \label{eq:amp2}
    \braket{\nu_{\mu}^{\mathcal{PT}}(t_0)|\nu_{e}(t)}&={\rm sech} 2\theta\,e^{-i\omega_j t_0+i\omega_k t}\nonumber\\&\phantom{=} \quad\times P_{kj}(V^{-1})_{j \mu}^*(V^{-1})_{e k}\,,
\end{align}
where, with a slight abuse of notation, we have used
\begin{equation}
    \braket{\nu_j^{\mathcal{PT}}(0)|\nu_k(0)}\equiv P_{kj}={\rm diag}(\pm 1,\mp 1)\,.
\end{equation}
(Note that $P$ here is not the parity matrix, but rather keeps track of which of the mass eigenstates has positive and which has negative norm with respect to $\mathcal{PT}$. The overall sign of $P$ is a matter of convention.) We reiterate that the inner products of different flavour states are being taken with respect to the $\mathcal{PT}$ inner product, such that the state space is being spanned either by $\{\ket{\nu_{e}},\ket{\nu_{\mu}^{C'}}\}$ or $\{\ket{\nu_{e}^{C'}},\ket{\nu_{\mu}}\}$ and, as before, the normalisation of the flavour states has been rescaled by a factor $\sqrt{{\rm sech} 2\theta}$.

For $V$ of the form in Eq.~\eqref{eq:Vmatrix}, the transition probability obtained from the product of Eqs.~\eqref{eq:amp1} and~\eqref{eq:amp2} is
\begin{equation}
    \mathbb{P}_{\nu_{\mu}\to\nu_{e}}(t,t_0)=\tanh^2 2\theta\sin^2[\Delta \omega \Delta t/2]\,,
\end{equation}
consistent with the earlier result, wherein the phase $\varphi$ has cancelled. We can consider the ultrarelativistic limit $\Delta \omega\to |\Delta \omega| \simeq (m_1^2-m_2^2)/(2E)$ and $\Delta t=L$ (for $c=1$) and recover the familiar two-neutrino oscillation probability with the $\sin^2(2\theta)$ mixing-angle factor that appears in the Hermitian case replaced by $\tanh^2(2\theta)$. This result is not simply the analytic continuation $\theta \to \pm i \theta$ of the Hermitian case. However, the range $[0,1]$ of the mixing-angle factor is the same in the two cases, so two-flavour mixing data can be analysed in the same way, although with a different interpretation in terms of neutrino mass parameters. Specifically, we see from Eq.~\eqref{eigenmasses} that the masses of the two neutrino species would become degenerate in the limit $\tanh^2 2\theta\ \to 1$, corresponding to an exceptional point that has no analogue in the Hermitian case.

In summary, the results of this Letter place the treatment of non-Hermitian flavour mixing matrices on a firm footing, laying the foundation for a consistent treatment of flavour oscillations and $CP$ violation in non-Hermitian extensions of the complete quark and lepton sectors of the Standard Model of particle physics.
Given the existing and expected constraints on quark mixing and neutrino oscillation phenomena, we re-emphasise that the construction of transition probabilities for the two-state, $\mathcal{PT}$-symmetric model described here is in principle experimentally testable. In particular, we have identified that the relationship between the mass splitting and strength of the mixing differs between the non-Hermitian and Hermitian cases. A full analysis of the degree of sensitivity of experiments and observations to this difference is beyond the scope of this work. In the case of neutrino physics, such an analysis may rely on the extension to three-neutrino mixing, and to other neutrino observables, which we leave for future work.\\


\section*{Acknowledgments}

This work was motivated in part by the Masters Thesis of MD and RM, supervised at the University of Nottingham by PM~\cite{MDRMThesis}. PM would like to thank King's College London for the support of visitor status. The work of JA and JE was supported by the United Kingdom Science and Technology Facilities Council (STFC) [Grant No.~ST/T000759/1] and Engineering and Physical Sciences Research Council (EPSRC) [Grant No.~EP/V002821/1]. The work of MD was supported by the European Union’s Horizon 2020 research and innovation programme under grant agreement No.~765048. The work of RM was supported by an EPSRC PhD studentship grant [Grant No.~EP/R513271/1] and the Deutsche Forschungsgemeinschaft (DFG, German Research Foundation) under grant 396021762 - TRR 257. The work of PM was supported by a Nottingham Research Fellowship from the University of Nottingham, the Science and Technology Facilities Council (STFC) [Grant No.~ST/X00077X/1], and a United Kingdom Research and Innovation (UKRI) Future Leaders Fellowship [Grant Nos.~MR/V021974/1 and~MR/V021974/2]. For the purpose of open access, the authors have applied a Creative Commons Attribution (CC BY) licence to any Author Accepted Manuscript version arising.

\section*{Data Access Statement}

The data that support the findings of this study are available upon reasonable request from the authors.



\section*{Supplemental Material}

\begin{figure}
    \centering
    \includegraphics[width=0.5\textwidth]{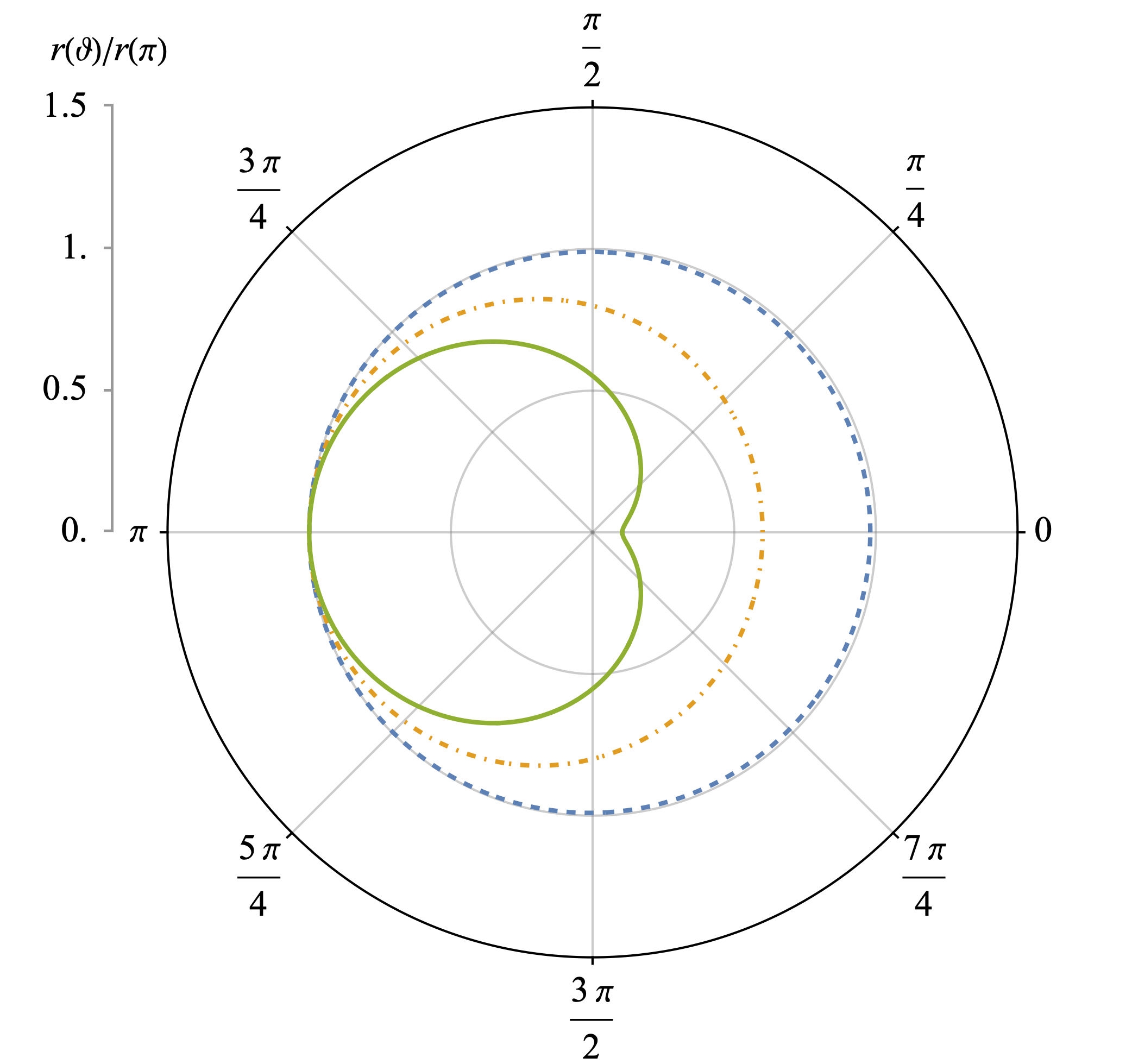}
    \caption{Polar plot of the Dirac norm $r(\vartheta)/r(\pi)$ for different values of the parameter $\eta$:\ that for $\eta=0.1$ (dashed, blue) is close to a unit circle, centered on the origin; that for
    $\eta=0.5$ (dot-dashed, yellow) clearly deviates from the unit circle; and that for $\eta=0.9$ (solid, green) has a distinctive cardioid shape.}
    \label{trajectories}
\end{figure}

We briefly illustrate in this Supplemental Material the violation of time-translation invariance that results if one attempts to construct transition probabilities with respect to the Dirac inner product. The Dirac-conjugate states
\begin{subequations}
\bea
\bra{\phi_1(t)}&=&\cosh(\theta)\,\xi_+^{\ast}(t)\,\mathbf{e}
_+^{\dag}+\sinh(\theta)\,\xi_-^{\ast}(t)\,\mathbf{e}_-^{\dag}\,,\ \\
\bra{\phi_2(t)}&=&\cosh(\theta)\,\xi_-^{\ast}(t)\,\mathbf{e}_-^{\dag}+\sinh(\theta)\,\xi_+^{\ast}(t)\,\mathbf{e}_+^{\dag}\,,\ 
\eea
\end{subequations}
lead to time-dependent norms
\be\label{norm}
\braket{\phi_1(t)|\phi_1(t)} = \braket{\phi_2(t)|\phi_2(t)}\ =\ \frac{1-\eta^2\cos(\Delta\omega t)}{1-\eta^2}\,.
\ee
The Dirac norm traces out a cardioid as shown in Fig.~\ref{trajectories}, wherein we have defined $r(\vartheta)=[1-\eta^2\cos(\vartheta)]/(1-\eta^2)$. The flavour states are orthogonal with respect to the Dirac inner product only at $t=0$:
\bea
&&\braket{\phi_1(t)|\phi_2(t)} = \braket{\phi_2(t)|\phi_1(t)}^{\ast}\ =\ \frac{\eta}{1-\eta^2}\nonumber\\&&\qquad \times \Big[1-\cos (\Delta \omega t)-i\sqrt{1-\eta^2}\sin(\Delta\omega t)\Big]\,,
\eea
and the violation of time-translation invariance is made manifest by any adjustment of the normalisation of the states, e.g., by defining
$\ket{\bar\phi_i(t)}\equiv\ket{\phi_i(t)}/\sqrt{\braket{\phi_i(t)|\phi_i(t)}}$.


\end{document}